\documentclass[preprint2,usenatbib,twocolappendix]{aastex631}
\usepackage{lipsum}
\usepackage{totcount}
\usepackage{comment}
\usepackage{amsmath}
\newtotcounter{citnum} \def\oldbibitem{} \let\oldbibitem=\bibitem
\def\bibitem{\stepcounter{citnum}\oldbibitem}
\usepackage{array}
\usepackage{floatrow}
\usepackage[caption=false]{subfig}
\graphicspath{{./}{figures/}}

\shorttitle{Detection of CrH in WASP-31b}
\shortauthors{Flagg et al.}

\begin{document}

\title{ExoGemS Detection of a Metal Hydride in an Exoplanet Atmosphere\\ at High Spectral Resolution}

\correspondingauthor{Laura Flagg}
\email{laura.s.flagg@gmail.com}

\author[0000-0001-6362-0571]{Laura Flagg}
\affil{Department of Astronomy and Carl Sagan Institute, Cornell University, Ithaca, New York 14853, USA}

\author[0000-0001-7836-1787]{Jake D. Turner}
\affil{Department of Astronomy and Carl Sagan Institute, Cornell University, Ithaca, New York 14853, USA}
\affil{NHFP Sagan Fellow}

\author[0000-0001-9796-2158]{Emily Deibert}
\affiliation{Gemini Observatory, NSF’s NOIRLab, Casilla 603, La Serena, Chile}

\author[0000-0002-5425-2655]{Andrew Ridden-Harper}
\affil{Las Cumbres Observatory, 6740 Cortona Drive, Suite 102, Goleta, CA 93117, USA}

\author[0000-0001-6391-9266]{Ernst de Mooij}
\affiliation{Astrophysics Research Centre, School of Mathematics and Physics, Queen’s University Belfast, University Road,
Belfast BT7 1NN, United Kingdom}

\author[0000-0003-4816-3469]{Ryan J. MacDonald}
\affil{Department of Astronomy, University of Michigan, Ann Arbor, MI 48109, USA}
\affil{Department of Astronomy and Carl Sagan Institute, Cornell University, Ithaca, New York 14853, USA}
\affil{NHFP Sagan Fellow}

\author[0000-0001-5349-6853]{Ray Jayawardhana}
\affil{Department of Astronomy, Cornell University, Ithaca, New York 14853, USA}

\author[0000-0002-9308-2353]{Neale Gibson}
\affil{School of Physics, Trinity College Dublin, The University of Dublin, Dublin 2, Ireland}

\author[0000-0002-4451-1705]{Adam Langeveld}
\affil{Department of Astronomy and Carl Sagan Institute, Cornell University, Ithaca, New York 14853, USA}

\author[0000-0001-6050-7645]{David Sing}
\affil{Department of Earth \& Planetary Sciences, Johns Hopkins University,  Baltimore, MD, 21218, USA}

\begin{abstract}
Exoplanet atmosphere studies are often enriched by synergies with brown dwarf analogs.  However, many key molecules commonly seen in brown dwarfs have yet to be confirmed in exoplanet atmospheres. An important example is chromium hydride (CrH), which is often used to probe atmospheric temperatures and classify brown dwarfs into spectral types. Recently, tentative evidence for CrH was reported in the low-resolution transmission spectrum of the hot Jupiter WASP-31b. Here, we present high spectral resolution observations of WASP-31b's transmission spectrum from GRACES/Gemini North and UVES/VLT. We detect CrH at 5.6$\sigma$ confidence, representing the first metal hydride detection in an exoplanet atmosphere at high spectral resolution. Our findings constitute a critical step in understanding the role of metal hydrides in exoplanet atmospheres.

\end{abstract}

\keywords{Exoplanet atmospheres (487) --- Planetary atmospheres (1244) 	--- Exoplanets (498) --- Exoplanet atmospheric composition (2021)}

\section{Introduction} \label{sec:intro}

Metal hydrides and metal oxides have been used to characterize the atmospheres of very cool stars and brown dwarfs for decades. TiO is useful for classifying stars as warm as $\sim$4000 K \citep{ReidPalomarMSUNearbyStar1995}, while metal hydrides like FeH and CrH become the primary tools for determining the temperatures of brown dwarfs with T$_{\rm{eff}}\lesssim$2500 K when using optical spectra \citep{KirkpatrickDwarfsCoolerDefinition1999, MartinSpectroscopicClassificationLateM1999}. CrH is also a sensitive probe of rainout chemistry and disequilibrium atmospheric dynamics in brown dwarfs \citep{BurrowsNewCrHOpacities2002}.

Many hot giant exoplanets share similar temperatures to brown dwarfs, so metal oxides and hydrides could also sculpt exoplanet spectra and drive thermal inversions \citep{FortneyUnifiedTheoryAtmospheres2008,GandhiNewAvenuesThermal2019}. However, searches for TiO and VO have had limited success, with initial detections often challenged by subsequent observations 
\citep[e.g.][]{NugrohoHighresolutionSpectroscopicDetection2017, HermanSearchTiOOptical2020}. There is no consensus on the reason for the lack of detections. Chemical equilibrium models indicate that, as with brown dwarfs, molecules such as CrH and FeH should have significant abundances in exoplanet atmospheres from $\sim$ 1200--2000 K \citep[e.g.][]{VisscherAtmosphericChemistryGiant2010}.

Observational evidence for metal hydrides in exoplanet atmospheres has been tentatively reported from low spectral resolution transmission spectra for a handful of planets. Evidence of FeH has been reported in WASP-79b's Hubble Space Telescope (HST) WFC3 transmission spectrum \citep{SotzenTransmissionSpectroscopyWASP79b2020,SkafARESIICharacterizing2020}, though subsequent observations with HST's STIS instrument favored unocculted stellar faculae with no FeH \citep{RathckeHSTPanCETProgram2021}. Hints of CrH have been inferred for the warm Neptune HAT-P-26b \citep{MacDonaldMetalrichAtmosphereExoNeptune2019} and the hot Jupiter WASP-31b \citep{BraamEvidenceChromiumHydride2021}. However, the low spectral resolution observations underpinning these inferences render the robust identification of a chemical species challenging, since bands from alternative molecules can often be hard to distinguish. 

High-resolution spectroscopy offers a promising avenue to search for metal hydrides in giant exoplanet atmospheres. Since absorption lines from individual molecules are  resolved at high resolution, specific molecules can be robustly and directly detected. To date, the only attempt to detect metal hydrides at high spectral resolution was by \citet{KesseliSearchFeHHotJupiter2020}, who surveyed 12 planets and reported no evidence for FeH. The definitive detection of metal hydrides would be an important advancement in our understanding of hot giant planet atmospheres, unveiling their role in driving the thermal structure of hot giant planet atmospheres.

In this letter, we report the detection of CrH at high spectral resolution in the atmosphere of \object[WASP-31]{WASP-31b}. We present high-resolution spectra from both the ExoGemS survey with Gemini-North/GRACES and VLT/UVES, both of which establish the presence of CrH with a combined significance of $>$5$\sigma$. This marks the first detection of a metal hydride from a high-resolution exoplanet spectrum. 

\section{WASP-31b}

\begin{deluxetable*}{lccc}
\tablecaption{Orbital and physical parameters of the WASP-31 system used in this analysis. \label{tab:parameters}}
\tablehead{\colhead{Parameter} & \colhead{Symbol (Unit)} & \colhead{Value} & \colhead{Reference}
    }
\startdata
Stellar mass & $M_*$ ($M_\sun$) & $1.15 \pm 0.08$ & \cite{AndersonWASP31bLowdensityPlanet2011}  \\
Projected stellar rotational velocity & $v\sin i$ (km/s) & 7.9 $\pm$ 0.6 & \cite{AndersonWASP31bLowdensityPlanet2011} \\
Magnitude & $V$ (mag) & $11.90 \pm 0.02$ & \cite{ZachariasFourthUSNaval2013} \\
Orbital period & $P$ (days) & 3.4058864$\pm$0.0000018  & this paper \\
Epoch of mid-transit & $t_c$ (BJD) & 2455873.8673$\pm$0.0017 & this paper \\
Transit duration & $T_{14}$ (hours) & $2.6472 \pm 0.0312 $ & \cite{AndersonWASP31bLowdensityPlanet2011} \\
Planetary radius & $R_\mathrm{p}$ ($R_\mathrm{J}$) & $1.549 \pm {0.050}$ & \cite{AndersonWASP31bLowdensityPlanet2011}  \\
Planetary mass & $M_\mathrm{p}$ ($M_\mathrm{J}$) & $0.478\pm 0.029$ & \cite{AndersonWASP31bLowdensityPlanet2011} \\
Equilibrium Temperature & $T_\mathrm{eq}$ (K) & 1393 & \cite{BraamEvidenceChromiumHydride2021} \\
Inclination & $i$ (degrees) & $84.41{}^{+0.22}_{-0.22}$ & \cite{AndersonWASP31bLowdensityPlanet2011} \\
Systemic velocity & $\gamma_\mathrm{sys}$ (km/s) & $-0.34 \pm 2.07$ & \cite{GaiaCollaborationGaiaDataRelease2018} \\
Stellar radial velocity semi-amplitude & $k_*$  (m/s) & $59.4{}^{+0.28}_{-0.29}$ & \cite{BonomoGAPSProgrammeHARPSN2017} \\
Planetary radial velocity semi-amplitude & $k_p$ (km/s) & $148.0 \pm 14.8$ &  this paper \\
Linear limb darkening coefficient & $u_1$ & 0.2387 & \cite{ClaretLimbGravitydarkeningCoefficients2017} \\
Quadratic limb darkening coefficient & $u_2$ & 0.3118 & \cite{ClaretLimbGravitydarkeningCoefficients2017} 
\enddata
\end{deluxetable*}

The hot Jupiter WASP-31b, discovered by \citet{AndersonWASP31bLowdensityPlanet2011}, orbits an F5 star on a 3.4 day orbit.  (See Table \ref{tab:parameters} for the full parameters of the system used in this paper). At 0.5 $M_\mathrm{J}$ and 1.5 $R_\mathrm{J}$ \citep{AndersonWASP31bLowdensityPlanet2011}, it is has extremely low-density, even for a giant planet; its equilibrium temperature is $\sim$1400 K \citep{BraamEvidenceChromiumHydride2021}.  \citet{BrownRossiterMcLaughlinEffectMeasurements2012} used the Rossiter-McLaughlin effect  to measure the alignment of the orbit with the rotation of \object{WASP-31}: they found $\lambda = 2.8 \pm 3.4$ degrees, implying the system is very well aligned.

\subsection{The Atmosphere of WASP-31b}

\citet{SingHSTHotJupiterTransmission2015} used HST and Spitzer to probe the atmosphere between 0.3 and 1.7 $\mu$m.  In the optical, they found no Na, but evidence of K at a 4.2$\sigma$ confidence level.  The expected H$_2$O feature in the infrared was also undetected, implying the presence of clouds.  Several papers have reanalyzed the HST data and reported fairly similar results \citep{BarstowConsistentRetrievalAnalysis2017,WelbanksMassMetallicityTrendsTransiting2019}. \citet{MacDonaldSignaturesNitrogenChemistry2017} noted weak evidence for NH$_3$ using HST-WFC3 data.

\citet{GibsonVLTFORS2Comparative2017} used VLT/FORS2 to characterize the atmosphere during transit.  They confirmed the cloud deck seen by \citet{SingHSTHotJupiterTransmission2015}, but saw no evidence for K.  \citet{GibsonRevisitingPotassiumFeature2019} followed up with observations using VLT/UVES, and did not find evidence of K either.  \citet{McGruderACCESSConfirmationNo2020} reported the same, using IMACS on Magellan.

\citet{BraamEvidenceChromiumHydride2021} reanalyzed the HST data from \citet{SingHSTHotJupiterTransmission2015}.  They found significant --- albeit inconclusive --- evidence for not only H$_2$O but also CrH, the latter of which has no confirmed detections in any exoplanet.

\section{Observations}
We observed one transit of WASP-31b on 2022-March-12 as part of the Exoplanets with Gemini Spectroscopy (ExoGemS) survey. ExoGemS uses  GRACES  (Gemini Remote Access to CFHT ESPaDOnS Spectrograph) on Gemini-North to observe a large sample of exoplanets with transmission spectroscopy. GRACES covers from 400 to 1050 nm at R$\sim$66,000.

\begin{figure}
\centering
\includegraphics[width=0.98\textwidth]{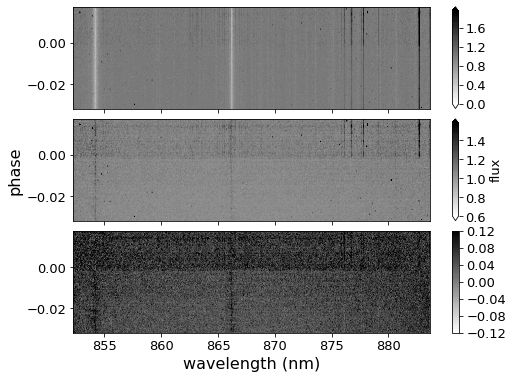}
\caption{One order of the reduced spectra from GRACES (top), with the main stellar and telluric signal divided out (middle), and after SVD processing (bottom). \label{fig_gracesspec}}
\end{figure}

141 spectra were acquired between 09:17 and 13:21 UTC, corresponding to airmasses between 1.28 and 1.94, for an expected transit between 10:36 and 13:11 UTC.  In the order with the most prominent CrH band, $\sim$860 nm, the SNR ranged from 32 to 85. There was a drop in SNR around the time of mid-transit, with spectra before mid-transit having an average SNR of 65 and spectra after having an average SNR of 41.  One order of the reduced spectra is plotted in the top panel of Figure \ref{fig_gracesspec}.

The GRACES data was supplemented by archival VLT/UVES data of two WASP-31b transits, taken in the spring of 2017 and published by \citet{GibsonRevisitingPotassiumFeature2019}. As that data were not intended to look for CrH, the chosen wavelength range ends at 866 nm and only overlaps with the strongest CrH band, which starts at 860 nm, for $\sim$6 nm.

\section{Methods}
\subsection{Calculation of the Transit Ephemeris}

To update the transit ephemeris we analyzed the publicly available TESS data. TESS observed WASP-31b in Sectors 9 (2019-02-28 to 2019-03-26, \dataset[doi:10.17909/yk3b-g272]{\doi{10.17909/yk3b-g272}}) and 36 (2021-03-07 to 2021-04-02, \dataset[doi:10.17909/x636-7g89]{\doi{10.17909/x636-7g89}}). We use the Data Validation Timeseries  light curves  because they have less scatter in their out-of-transit (OoT) baseline.  We followed the same procedure as in \citet{TurnerDecayingOrbitHot2021,TurnerCharacterizingWASP4System2022} to model the light curves and determine the transit ephemeris. 

We modeled the TESS transits of WASP-31b with the EXOplanet MOdeling Package (\texttt{EXOMOP}; \citealt{PearsonPhotometricObservationHATP16b2014,TurnerInvestigatingPhysicalProperties2017})\footnote{\texttt{EXOMOPv7.0}; \href{https://github.com/astrojake/EXOMOP}{https://github.com/astrojake/EXOMOP} } to find a best-fit. \texttt{EXOMOP} creates a model transit using the analytic equations of \cite{MandelAnalyticLightCurves2002} and the data are modeled using a Differential Evolution Markov Chain Monte Carlo (DE-MCMC; \citealt{EastmanEXOFASTFastExoplanetary2013}) analysis and uses the residual permutation, time-averaging, and wavelet methods to account for red noise. Each TESS transit was modeled with \texttt{EXOMOP} independently. We used 20$^{6}$ links and 20 chains for the DE-MCMC model and use the Gelman-Rubin statistic (\citealt{GelmanInferenceIterativeSimulation1992}) to ensure chain convergence (\citealt{FordImprovingEfficiencyMarkov2006}). The mid-transit time ($t_{c}$), scaled semi-major axis ($a/R_{*}$), planet-to-star radius ($R_{p }/R_{*}$), and inclination ($i$) are set as free parameters for every transit. The linear and quadratic limb darkening coefficients and period are fixed during the analysis. The linear and quadratic limb darkening coefficients are taken from \citet{ClaretLimbGravitydarkeningCoefficients2017} and are set to 0.2387 and 0.3118, respectively. We find an updated period of 3.4058864$\pm$0.0000018 days and a mid-transit (t$_{c}$[0]) of 2455873.8673$\pm$0.0017 BJD$_{TDB}$ using the TESS observations (see Table \ref{tb:TESS_mid} for the individual TESS mid-transit times).

\subsection{Removal of Stellar and Telluric Signal}
As in \citet{DeibertDetectionIonizedCalcium2021}, the ExoGemS spectra were reduced with the OPERA pipeline \citep{MartioliOpenSourcePipeline2012},  designed for ESPaDOnS data, which extracts the spectra, calibrates the wavelength solution, and   removes the blaze function.  The UVES spectra were reduced in the manner of \citet{GibsonRevisitingPotassiumFeature2019}.  After the spectral extraction, we used the same procedure for both data sets for consistency.

\begin{figure}[ht!]
\centering
\includegraphics[width=0.98\textwidth]{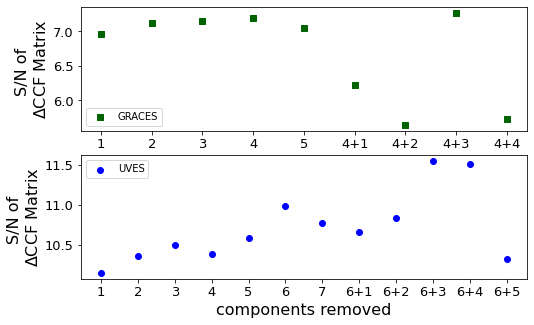}
\caption{The S/N of the injected signal in the $\Delta$CCF matrix as a function of components removed for GRACES (top) and UVES (bottom).  The plus signal indicates additional components removed after the barycentric correction to better remove the stellar signal.\label{fig:deltaccf}}
\end{figure}

The basic procedure for removing the stellar and telluric features relies on the fact that they are essentially stationary in time, while the planet's signal is moving in Doppler space \citep[e.g][]{ BrogiDetectionMolecularAbsorption2013, LockwoodNearIRDirectDetection2014, FlaggCODetectedCI2019}.  For example, over the course of the GRACES observations, the planet's signal shifts 45 km/s. The specific procedure used for this paper is a custom code.\footnote{\href{https://github.com/lauraflagg/svd\_exoplanets}{https://github.com/lauraflagg/svd\_exoplanets}} We first put all spectra onto a common wavelength grid that is evenly spaced in velocity. We then divided by the median out-of-transit spectrum to remove the blaze function and the basic spectral features.  We further  removed the stellar and telluric features using  singular value decomposition (SVD) as in \citet{LineSolarSubsolarMetallicity2021}.  For GRACES, we used 4 components; for UVES we used 6.  The number of components was chosen using the $\Delta$CCF method \citep[e.g.][]{SpringBlackMirrorImpact2022, CheverallRobustnessMeasuresMolecular2023}: we choose the number of components that optimizes the signal-to-noise (S/N) of an injected signal in the CCF matrix after subtracting the original CCF matrix (Figure \ref{fig:deltaccf}). We then weight the data by the variance in each wavelength bin for a given night.  The data from different orders are then combined into a 1D spectrum, so we can leverage the ratio of line strengths between orders. 

  We then corrected for barycentric shift in each spectrum, as calculated with \textit{astropy.time}, so that all the stellar signals would be aligned.  We then removed any remaining signals from the star, using a second application of SVD, removing  3 components for GRACES and both UVES transits (bottom panel, figure \ref{fig_gracesspec}). This is needed in wavelength regions with strong stellar lines, because the stellar signal moves by over $\sim$500 m/s due to the barycentric shift during our observations.  The improvement is seen in the recovery of the injected signal in the $\Delta$CCF matrix (Figure \ref{fig:deltaccf}).
  
  We do not use the entire wavelength range in our final calculations.    At many wavelengths, a given molecule may have approximately no opacity, so we would not expect to detect it even at much higher SNR.  Other wavelengths have significant contamination from incomplete removal of telluric and/or stellar features.  The worst stellar contamination, which in this case is around the Ca II Infrared triplet lines, are masked out, as are the O$_2$ telluric bands.  For CrH, we concentrated on  a bandhead at 860 nm.  For both GRACES and UVES, the strongest lines are all in a single order.  For UVES, the spectra cut off around 866 nm, so we simply use that as the endpoint.  For GRACES, the data in principle goes past 950 nm, well into the H$_2$O telluric band. As the CrH features also get weaker as you go to longer wavelengths, we expect there would be  very  little signal added in a second order.  Nevertheless, we did try the wavelength range from the second order (886 to 919 nm) both on its own and as one longer spectra from 860 to 919 nm.  As the signal strength did not vary significantly with endpoints between 885 and 919 nm, we tried using different endpoints and chose the endpoint that maximized any detection significance.  In this case, the final wavelength range for the GRACES data was 860 nm to 895 nm.

\subsection{Deriving the Transmission Spectrum}
We use standard methods for creating the CCF matrix to evaluate the presence of molecules in exoplanet atmospheres \citep[e.g.][]{BirkbyDetectionWaterAbsorption2013, RodlerDetectionCOAbsorption2013}.  We first shift all spectra into the planetary rest frame using the equation:
\begin{equation}
    \Delta v=k_p\sin\left(\frac{2\pi}{P}(t-t_c)\right)
\end{equation}
 where $t$ is the date, $t_c$ is the date of the transit midpoint,  $P$ is the period, $\Delta v$ is the velocity shift, and $k_p$ is the velocity amplitude of the planet.  While in principle $k_p$ can be calculated from stellar and planet orbital parameters, we chose to allow $k_p$ to be a free parameter which we allow to vary between -200 and 200 km/s.  Negative values for $k_p$, while physically unrealistic, allow us to better determine if any signals we see are false positives.  The systemic velocities were chosen based on the initial pixel spacing for each spectrograph, so we sample the UVES data more finely than the GRACES data.   Once shifted,  we cross-correlate the spectra with the template before coadding the CCFs. This procedure was implemented using a custom code.\footnote{\href{https://github.com/lauraflagg/combine-and-xcor}{https://github.com/lauraflagg/combine-and-xcor}}.

\subsection{CrH Model Templates}

\begin{figure}[ht!]
\centering
\includegraphics[width=0.98\textwidth]{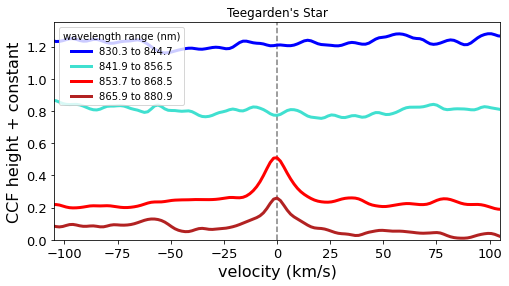}
\caption{Cross-correlating our template with the spectra of a very low-mass star yields a clear signal in orders where CrH is expected to be present (red lines, covering wavelengths longer than 860 nm) and not the orders where it would not be present (blue lines).  \label{fig_crhbd}}
\end{figure}

We created model CrH transmission spectra templates using the TRIDENT radiative transfer code \citep{MacDonaldTRIDENTRapid3D2022}, which employs the CrH line list from \citet{BurrowsNewCrHOpacities2002}. Since this is an old line list, we validated it by cross-correlating our template with a high-resolution, optical spectrum of a very low-mass, cold star (Teegarden's Star, M8) from CARMENES DR1 \citep{RibasCARMENESSearchExoplanets2023}. Very low-mass stars and warmer brown dwarfs exhibit CrH absorption starting at 860 nm \citep[e.g.][]{KirkpatrickDwarfsCoolerDefinition1999}.   In Figure \ref{fig_crhbd}, we plot these CCFs,  which show a clear detection of CrH in the orders where we would expect to see CrH in a very low-mass star or brown dwarf.

\begin{figure*}[ht!]
\centering
\includegraphics[width=0.98\textwidth]{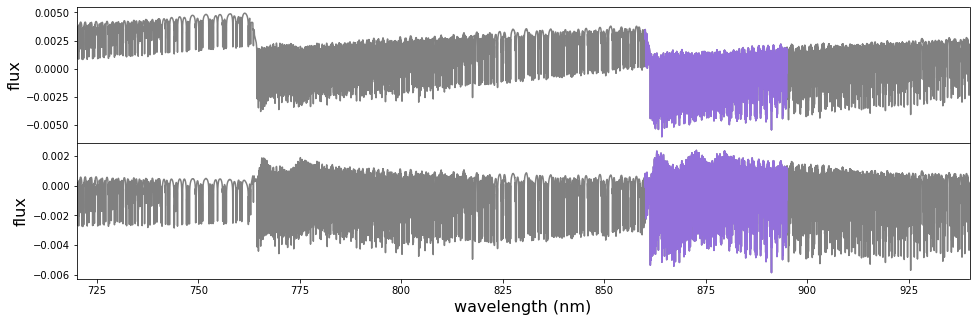}
\caption{The original (top) and filtered CrH template (bottom). The wavelength range use, chosen because of the relatively strong CrH lines with minimal telluric contamination, is highlighted in purple.  \label{fig_template}}
\end{figure*}

We next created a grid of templates with isothermal temperatures from 1000 to 2200\,K in increments of 400 K and volume mixing ratios (VMR) from $\log_{10}$(VMR) = -6 to -10 in increments of 1 dex. Each model atmosphere consists of 100 layers spaced uniformly in log-pressure from $10^{-9}$ to $100$\,bar. The original high-resolution model transmission spectra were computed at $\Delta\nu = 0.01$\,cm$^{-1}$ ($R = 10^6$ at 1\,$\micron$) and subsequently resampled to the respective resolutions of the two spectrographs, resulting in two sets of grids. We first multiplied the templates by -1, so that a detection of atmospheric absorption results in a positive correlation. We also subtract the mean and apply a Butterworth Filter to the template, as in  \cite{HermanDaysideFeEmission2022}, to mimic the effect of the SVD on the template itself. Figure~\ref{fig_template} shows the original version and the filtered version of the best-fit CrH template for the GRACES data. We tried using solely the 860 nm CrH band, as well as both the 764 nm and the 860 nm bands.  However, as the 764 band has significant telluric contamination, we used only the 860 nm band in the final analysis (Figure \ref{fig_template}).

\section{Results}
\subsection{CCF Matrix}\label{sec_ccf}

\begin{figure*}
\centering
\includegraphics[width=0.98\textwidth]{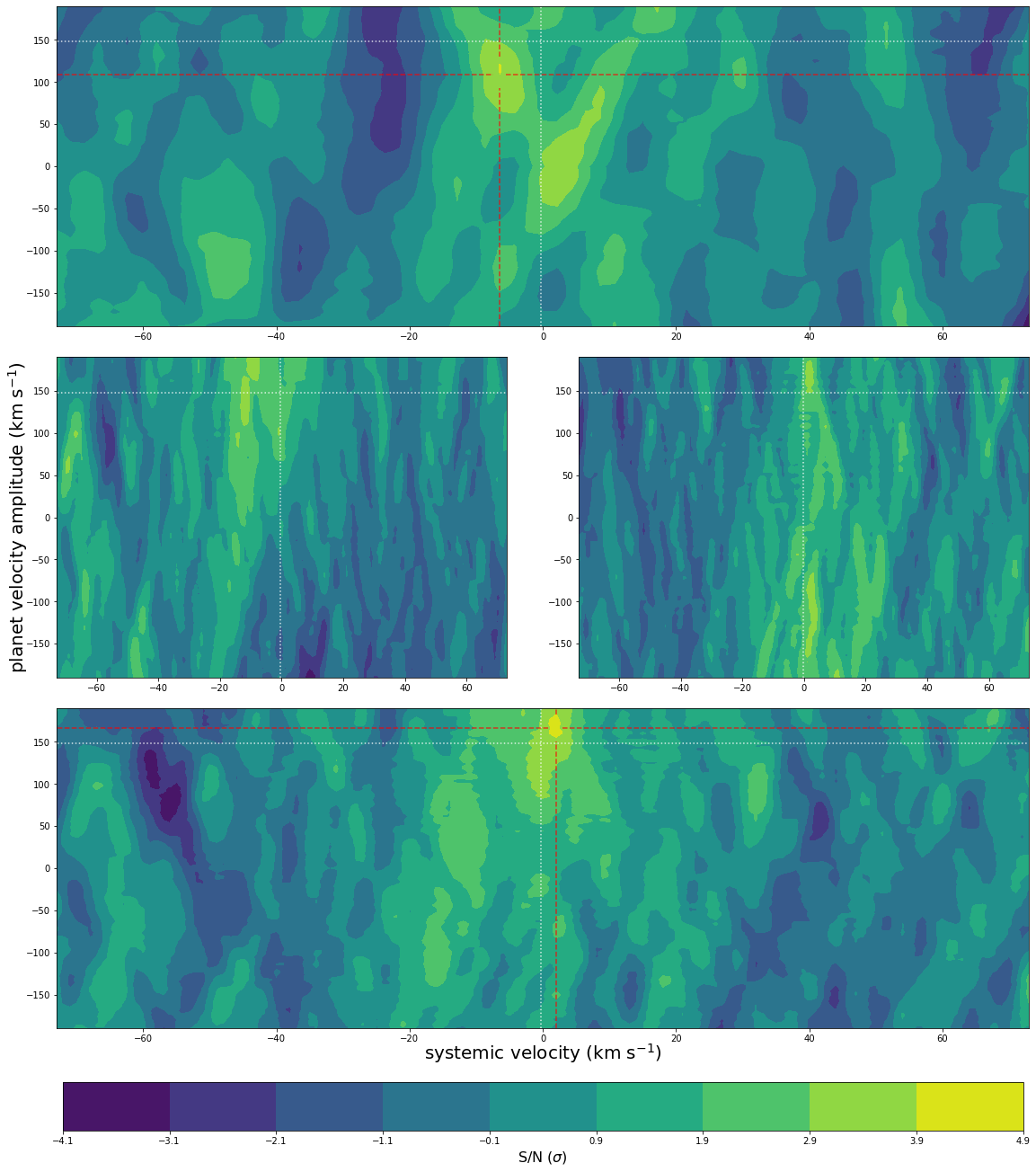}
\caption{The CCF matrices from GRACES (top), the 1st transit with UVES (middle, left),  the 2nd transit with UVES (middle, right), and both UVES transits combined (bottom).  For the three transits, we detect CrH at a 3.9, 3.7, and 3.6$\sigma$ level, respectively, while CrH is detected at a 4.2$\sigma$ level for the UVES combined data set.  The white dashed lines, indicating the systemic velocity of the system and the expected velocity amplitude of the planet (Table \ref{tab:parameters}), cross at the expected location of the planet's signal.  The red lines indicate the peaks' center positions. }
\label{fig_ccfmatrix}
\end{figure*}

  We initially tried the template with grid values closest to those in  \citet{BraamEvidenceChromiumHydride2021}, i.e., for an isothermal temperature of 1400 K and a VMR of 10$^{-8}$.   In the GRACES data, we find evidence of CrH at a 3.9$\sigma$ level (Figure \ref{fig_ccfmatrix}, top), consistent with the results from \citet{BraamEvidenceChromiumHydride2021}.   Our S/N is calculated by dividing the strength of the CCF by the standard deviation of the CCF matrix away from the detection.  The two UVES transits (Figure \ref{fig_ccfmatrix}, middle and bottom) independently confirm evidence of CrH at 3.7$\sigma$ and 3.6$\sigma$, respectively.   Combining both UVES transits yields a 4.2$\sigma$ CrH detection.

 All other templates with different CrH abundances or atmospheric temperatures yielded similar results. We find peak strengths within 1$\sigma$ of the default parameters in the CCF matrix, likely because the template does not change enough to impact the CCF. This is especially true given that the data are noisy. Consequently, our final analysis proceeded with the initial template parameters.

The location of the peak is not exactly the same for all three nights. This is not uncommon as previous studies at high resolution have seen similar discrepancies in the peak location \citep[e.g.][]{MerrittInventoryAtomicSpecies2021,Bello-ArufeMiningUltrahotSkies2022, Sanchez-LopezSearchingOriginEhrenreich2022}.  For a symmetric planet with no signatures of winds, we expect the peak to be at $k_p$=148.0 km/s and $v_{sys}$=-0.3 km/s. At that location, the GRACES data has dropped by $\sim$2.3$\sigma$ while the UVES data has dropped by 0.8$\sigma$.  These are large enough differences --- especially for the GRACES data --- that  random noise is unlikely to be the only cause.   Possible additional reasons include: 1) systematic issues, which we address with out-of-transit random sampling in Section \ref{sec_oot}; 2) varying SNR through the night, which will result in an offset if the planet is not completely symmetric;  3) an incorrect ephemeris, which can result in shifting signals by several km/s in $v_{sys}$ and 10s of km/s in $k_p$ (Meziani et al., in prep); or 4) an asymmetric planetary signal induced because the planetary atmosphere itself is not symmetric.  Since we do not know which cause is responsible, we cannot correct for it. 

\subsection{Bootstrap Sampling of Out-of Transit Spectra}\label{sec_oot}

To confirm the significance of our detection, we evaluated the CCF signal in another manner.  We use OoT spectra to model the systematic and random noise as in \citet{EstevesSearchWaterSuperEarth2017, DeibertHighresolutionTransitSpectroscopy2019}.   We did this twice, once for the GRACES data and once for both UVES transits.   For each of 20,000 iterations, we randomly chose the same number of OoT transits frames to correspond with each in-transit phase (89 for GRACES and 154 total between the two transits for UVES).  We  treated that set of spectra as if they were the observed spectra and created a simulated transit spectrum and then a CCF at $k_p$=148.0 km/s, the expected  $k_p$ value for this planet (Table \ref{tab:parameters}).  We then compared the simulated CCFs to the actual CCF. If the signal we detected in Section \ref{sec_ccf} is real, then it should be stronger than the vast majority of our simulated CCFs.   Because of the uncertainty on $v_{sys}$ plus the fact we fixed $k_p$, we allowed $v_{sys}$ to be slightly offset from the measured $v_{sys}$.

\begin{figure*}
\centering
\includegraphics[width=0.98\textwidth]{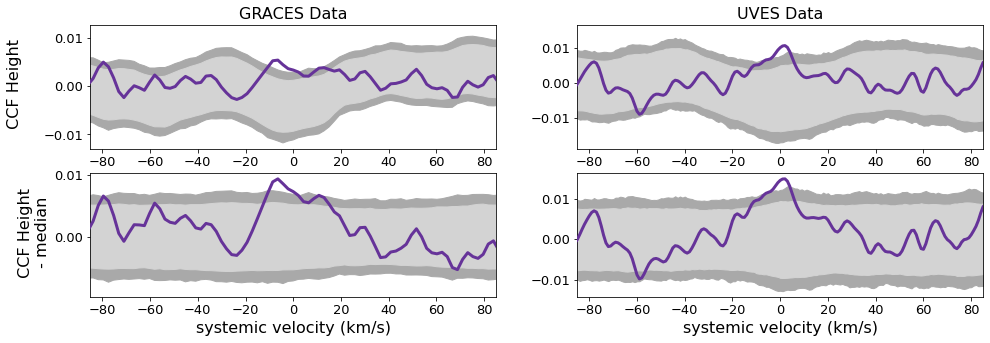}
\caption{The CCFs at the predicted $k_p$ (in purple) for WASP-31b along with the 99\% and 99.9\% ranges for the bootstrap sampling, with GRACES on the left and UVES on the right.    On the top row, we plot the raw CCF values, which shows the nature of the systematics for each instrument, while the bottom row shows the CCF values with the median from the bootstrap sample at each specific $v_{sys}$ subtracted.  Assuming a Gaussian distribution, these are equivalent to 3.9 and 3.5$\sigma$ detection respectively on their own, and a  5.6$\sigma$ detection when the probabilities are combined. \label{fig_oot}}
\end{figure*}

We plot the results in Figure \ref{fig_oot}; the left column is for GRACES data, while the right column is for UVES data.  In all four panels, the purple line is the actual CCF at the predicted $k_p$.  The shaded regions represent the 99\% region and the 99.9\% region.  On the top row, we plot the raw CCF values, which shows the nature of the systematics for each instrument, while the bottom row shows the CCF values with the median from the bootstrap sample at each specific $v_{sys}$ subtracted.  The GRACES maximum is above all but 99.995 \% of its bootstrap sample, while the UVES maximum is above 99.97 \% of its bootstrap sample.  Assuming a Gaussian distribution, we used \textit{scipy.stats.norm.ppf} to convert the one-sided probabilities into z-scores, resulting in values of $\sim$3.9 and  $\sim$3.5$\sigma$ significance, respectively. While not identical to the numbers we calculated in Section \ref{sec_ccf}, both are well within 1$\sigma$, showing that systematics from the spectrographs likely did not contribute much to the signals. Since these data sets can be assumed to be independent (i.e., taken on different dates, at different barycentric shifts, and with different instruments, with the caveat that unknown astrophysical systematics could create correlated noise in both data sets), we can simply multiply the probabilities to get a combined probability of 1 in $\sim$80 million.  Assuming a Gaussian distribution, this would be equivalent to a $\sim$5.6$\sigma$ detection, confirming that we see a real signal of CrH in WASP-31b's atmosphere.

\section{Conclusions}

Our detection of CrH --- along with the 3.3$\sigma$ evidence reported by  \citet{BraamEvidenceChromiumHydride2021} at low spectral resolution --- makes WASP-31b the first exoplanet with confirmed evidence of not just CrH, but any metal hydride. Moreover, our study represents the first detection of a metal hydride at high spectral resolution in an exoplanet atmosphere. 

Our results highlight a promising new window into hot Jupiter atmospheres offered by high-resolution observations. Spectral signatures of molecules like CrH are far more distinct and robust at high-resolution than at low-resolution, especially those with a low SNR. Basic templates can be created readily, so  quick checks for less commonly considered species like CrH are straightforward and more computationally feasible than running a full retrieval, as is typically done for low-resolution transit spectra. As seen with UVES data in this paper and in \citet{GibsonRevisitingPotassiumFeature2019}, high-resolution data can easily distinguish between potassium and CrH, even though CrH has a bandhead only 2 nm from one of the most prominent potassium features.

While WASP-31b is the first exoplanet with a confirmed detection of CrH, it will undoubtedly not be the last.  Detecting CrH in a single planet is merely the first step in potentially using CrH to characterize exoplanet atmospheres in a similar manner to brown dwarfs. While we are close to the limit of what is possible for characterizing CrH with ground-based, high-resolution spectroscopy today, CrH should be a viable target for exoplanets observed with JWST and the next generation of ground-based telescopes.

\section{Acknowledgments}

Thanks to the anonymous referee for their helpful comments.  We would also like to thank the very helpful team at Gemini-North, particularly Teo Mocnik, for their assistance preparing and acquiring the observations. RJ acknowledges support of a Rockefeller Foundation Bellagio Center residency.  

This work was enabled by observations made from the Gemini North telescope, located within the Maunakea Science Reserve and adjacent to the summit of Maunakea. We are grateful for the privilege of observing the universe from a place that is unique in both its astronomical quality and its cultural significance.

This research has made use of the VizieR catalogue access tool, CDS, Strasbourg, France. The original description of the VizieR service was published by  \cite{WengerSIMBADastronomicaldatabase2000}.  This research has made use of the SIMBAD database, operated at CDS, Strasbourg, France.  This research has made use of the Planetary Systems Composite Parameters Table at the NASA Exoplanet Archive \citep{NEA13}, which is operated by the California Institute of Technology, under contract with the National Aeronautics and Space Administration under the Exoplanet Exploration Program.

\facility{Gemini:Gillett, VLT:Kueyen, TESS, Exoplanet Archive}

\software{SpecTres \citep{CarnallSpectResFastSpectral2017}, NumPy \citep{oliphant2006guide, van2011numpy},  Pandas \citep{reback2020pandas},  Matplotlib \citep{Hunter:2007}, astropy \citep{CollaborationAstropyCommunityPython2013}, scipy \citep{2020SciPy-NMeth}}

\newpage
\clearpage 
\bibliographystyle{aasjournal} 

\begin{thebibliography}{}
\expandafter\ifx\csname natexlab\endcsname\relax\def\natexlab#1{#1}\fi
\providecommand{\url}[1]{\href{#1}{#1}}
\providecommand{\dodoi}[1]{doi:~\href{http://doi.org/#1}{\nolinkurl{#1}}}
\providecommand{\doeprint}[1]{\href{http://ascl.net/#1}{\nolinkurl{http://ascl.net/#1}}}
\providecommand{\doarXiv}[1]{\href{https://arxiv.org/abs/#1}{\nolinkurl{https://arxiv.org/abs/#1}}}

\bibitem[{Anderson {et~al.}(2011)Anderson, Collier~Cameron, Hellier, Lendl,
  Lister, Maxted, Queloz, Smalley, Smith, Triaud, West, Brown, Gillon, Pepe,
  Pollacco, S{\'e}gransan, Street, \&
  Udry}]{AndersonWASP31bLowdensityPlanet2011}
Anderson, D.~R., Collier~Cameron, A., Hellier, C., {et~al.} 2011, Astronomy and
  Astrophysics, 531, A60

\bibitem[{Barstow {et~al.}(2017)Barstow, Aigrain, Irwin, \&
  Sing}]{BarstowConsistentRetrievalAnalysis2017}
Barstow, J.~K., Aigrain, S., Irwin, P. G.~J., \& Sing, D.~K. 2017, The
  Astrophysical Journal, 834, 50

\bibitem[{{Bello-Arufe} {et~al.}(2022){Bello-Arufe}, Cabot, Mendon{\c c}a,
  Buchhave, \& Rathcke}]{Bello-ArufeMiningUltrahotSkies2022}
{Bello-Arufe}, A., Cabot, S. H.~C., Mendon{\c c}a, J.~M., Buchhave, L.~A., \&
  Rathcke, A.~D. 2022, The Astronomical Journal, 163, 96

\bibitem[{Birkby {et~al.}(2013)Birkby, {de Kok}, Brogi, {de Mooij}, Schwarz,
  Albrecht, \& Snellen}]{BirkbyDetectionWaterAbsorption2013}
Birkby, J.~L., {de Kok}, R.~J., Brogi, M., {et~al.} 2013, Monthly Notices of
  the Royal Astronomical Society, 436, L35

\bibitem[{Bonomo {et~al.}(2017)Bonomo, Desidera, Benatti, Borsa, Crespi,
  Damasso, Lanza, Sozzetti, Lodato, Marzari, Boccato, Claudi, Cosentino,
  Covino, Gratton, Maggio, Micela, Molinari, Pagano, Piotto, Poretti,
  Smareglia, Affer, Biazzo, Bignamini, Esposito, Giacobbe, H{\'e}brard,
  Malavolta, Maldonado, Mancini, Martinez~Fiorenzano, Masiero, Nascimbeni,
  Pedani, Rainer, \& Scandariato}]{BonomoGAPSProgrammeHARPSN2017}
Bonomo, A.~S., Desidera, S., Benatti, S., {et~al.} 2017, Astronomy and
  Astrophysics, 602, A107

\bibitem[{Braam {et~al.}(2021)Braam, {van der Tak}, Chubb, \&
  Min}]{BraamEvidenceChromiumHydride2021}
Braam, M., {van der Tak}, F. F.~S., Chubb, K.~L., \& Min, M. 2021, Astronomy
  and Astrophysics, 646, A17

\bibitem[{Brogi {et~al.}(2013)Brogi, Snellen, {de Kok}, Albrecht, Birkby, \&
  {de Mooij}}]{BrogiDetectionMolecularAbsorption2013}
Brogi, M., Snellen, I. A.~G., {de Kok}, R.~J., {et~al.} 2013, The Astrophysical
  Journal, 767, 27

\bibitem[{Brown {et~al.}(2012)Brown, Cameron, Anderson, Enoch, Hellier, Maxted,
  Miller, Pollacco, Queloz, Simpson, Smalley, Triaud, Boisse, Bouchy, Gillon,
  \& H{\'e}brard}]{BrownRossiterMcLaughlinEffectMeasurements2012}
Brown, D. J.~A., Cameron, A.~C., Anderson, D.~R., {et~al.} 2012, Monthly
  Notices of the Royal Astronomical Society, 423, 1503

\bibitem[{Burrows {et~al.}(2002)Burrows, Ram, Bernath, Sharp, \&
  Milsom}]{BurrowsNewCrHOpacities2002}
Burrows, A., Ram, R.~S., Bernath, P., Sharp, C.~M., \& Milsom, J.~A. 2002, The
  Astrophysical Journal, 577, 986

\bibitem[{Carnall(2017)}]{CarnallSpectResFastSpectral2017}
Carnall, A.~C. 2017, arXiv:1705.05165 [astro-ph].
\newblock \doeprint{1705.05165}

\bibitem[{Cheverall {et~al.}(2023)Cheverall, Madhusudhan, \&
  Holmberg}]{CheverallRobustnessMeasuresMolecular2023}
Cheverall, C.~J., Madhusudhan, N., \& Holmberg, M. 2023, Monthly Notices of the
  Royal Astronomical Society, 522, 661

\bibitem[{Claret(2017)}]{ClaretLimbGravitydarkeningCoefficients2017}
Claret, A. 2017, Astronomy \& Astrophysics, 600, A30

\bibitem[{Collaboration {et~al.}(2013)Collaboration, Robitaille, Tollerud,
  Greenfield, Droettboom, Bray, Aldcroft, Davis, Ginsburg, {Price-Whelan},
  Kerzendorf, Conley, Crighton, Barbary, Muna, Ferguson, Grollier, Parikh,
  Nair, Unther, Deil, Woillez, Conseil, Kramer, Turner, Singer, Fox, Weaver,
  Zabalza, Edwards, Azalee~Bostroem, Burke, Casey, Crawford, Dencheva, Ely,
  Jenness, Labrie, Lim, Pierfederici, Pontzen, Ptak, Refsdal, Servillat, \&
  Streicher}]{CollaborationAstropyCommunityPython2013}
Collaboration, A., Robitaille, T.~P., Tollerud, E.~J., {et~al.} 2013, Astronomy
  and Astrophysics, 558

\bibitem[{Deibert {et~al.}(2019)Deibert, {de Mooij}, Jayawardhana, Fortney,
  Brogi, Rustamkulov, \& Tamura}]{DeibertHighresolutionTransitSpectroscopy2019}
Deibert, E.~K., {de Mooij}, E. J.~W., Jayawardhana, R., {et~al.} 2019, The
  Astronomical Journal, 157, 58

\bibitem[{Deibert {et~al.}(2021)Deibert, {de Mooij}, Jayawardhana, Turner,
  {Ridden-Harper}, Fossati, Hood, Fortney, Flagg, MacDonald, Allart, \&
  Sing}]{DeibertDetectionIonizedCalcium2021}
---. 2021, The Astrophysical Journal, 919, L15

\bibitem[{Eastman {et~al.}(2013)Eastman, Gaudi, \&
  Agol}]{EastmanEXOFASTFastExoplanetary2013}
Eastman, J., Gaudi, B.~S., \& Agol, E. 2013, Publications of the Astronomical
  Society of the Pacific, 125, 83

\bibitem[{Esteves {et~al.}(2017)Esteves, {de Mooij}, Jayawardhana, Watson, \&
  {de Kok}}]{EstevesSearchWaterSuperEarth2017}
Esteves, L.~J., {de Mooij}, E. J.~W., Jayawardhana, R., Watson, C., \& {de
  Kok}, R. 2017, The Astronomical Journal, 153, 268

\bibitem[{Flagg {et~al.}(2019)Flagg, {Johns-Krull}, Nofi, Llama, Prato,
  Sullivan, Jaffe, \& Mace}]{FlaggCODetectedCI2019}
Flagg, L., {Johns-Krull}, C.~M., Nofi, L., {et~al.} 2019, The Astrophysical
  Journal Letters, 878, L37

\bibitem[{Ford(2006)}]{FordImprovingEfficiencyMarkov2006}
Ford, E.~B. 2006, The Astrophysical Journal, 642, 505

\bibitem[{Fortney {et~al.}(2008)Fortney, Lodders, Marley, \&
  Freedman}]{FortneyUnifiedTheoryAtmospheres2008}
Fortney, J.~J., Lodders, K., Marley, M.~S., \& Freedman, R.~S. 2008, The
  Astrophysical Journal, 678, 1419

\bibitem[{{Gaia Collaboration} {et~al.}(2018){Gaia Collaboration}, Brown,
  Vallenari, Prusti, {de Bruijne}, Babusiaux, {Bailer-Jones}, Biermann, Evans,
  Eyer, Jansen, Jordi, Klioner, Lammers, Lindegren, Luri, Mignard, Panem,
  Pourbaix, Randich, Sartoretti, Siddiqui, Soubiran, {van Leeuwen}, Walton,
  Arenou, Bastian, Cropper, Drimmel, Katz, Lattanzi, Bakker, Cacciari,
  Casta{\~n}eda, Chaoul, Cheek, De~Angeli, Fabricius, Guerra, Holl, Masana,
  Messineo, Mowlavi, Nienartowicz, Panuzzo, Portell, Riello, Seabroke, Tanga,
  Th{\'e}venin, {Gracia-Abril}, Comoretto, {Garcia-Reinaldos}, Teyssier,
  Altmann, Andrae, Audard, {Bellas-Velidis}, Benson, Berthier, Blomme, Burgess,
  Busso, Carry, Cellino, Clementini, Clotet, Creevey, Davidson, De~Ridder,
  Delchambre, Dell'Oro, Ducourant, {Fern{\'a}ndez-Hern{\'a}ndez}, Fouesneau,
  Fr{\'e}mat, Galluccio, {Garc{\'i}a-Torres}, {Gonz{\'a}lez-N{\'u}{\~n}ez},
  {Gonz{\'a}lez-Vidal}, Gosset, Guy, Halbwachs, Hambly, Harrison,
  Hern{\'a}ndez, Hestroffer, Hodgkin, Hutton, Jasniewicz,
  {Jean-Antoine-Piccolo}, Jordan, Korn, {Krone-Martins}, Lanzafame, Lebzelter,
  L{\"o}ffler, Manteiga, Marrese, {Mart{\'i}n-Fleitas}, Moitinho, Mora,
  Muinonen, Osinde, Pancino, Pauwels, Petit, {Recio-Blanco}, Richards,
  Rimoldini, Robin, Sarro, Siopis, Smith, Sozzetti, S{\"u}veges, Torra, {van
  Reeven}, Abbas, Abreu~Aramburu, Accart, Aerts, Altavilla, {\'A}lvarez,
  Alvarez, Alves, Anderson, Andrei, Anglada~Varela, Antiche, Antoja, Arcay,
  Astraatmadja, Bach, Baker, {Balaguer-N{\'u}{\~n}ez}, Balm, Barache, Barata,
  Barbato, Barblan, Barklem, Barrado, Barros, Barstow,
  Bartholom{\'e}~Mu{\~n}oz, Bassilana, Becciani, Bellazzini, Berihuete,
  Bertone, Bianchi, Bienaym{\'e}, {Blanco-Cuaresma}, Boch, Boeche, Bombrun,
  Borrachero, Bossini, Bouquillon, Bourda, Bragaglia, Bramante, Breddels,
  Bressan, Brouillet, Br{\"u}semeister, Brugaletta, Bucciarelli, Burlacu,
  Busonero, Butkevich, Buzzi, Caffau, Cancelliere, Cannizzaro, {Cantat-Gaudin},
  Carballo, Carlucci, Carrasco, Casamiquela, Castellani, {Castro-Ginard},
  Charlot, Chemin, Chiavassa, Cocozza, Costigan, Cowell, Crifo, Crosta,
  Crowley, Cuypers, Dafonte, Damerdji, Dapergolas, David, David, {de Laverny},
  De~Luise, De~March, {de Martino}, {de Souza}, {de Torres}, Debosscher, {del
  Pozo}, Delbo, Delgado, Delgado, Di~Matteo, Diakite, Diener, Distefano,
  Dolding, Drazinos, Dur{\'a}n, Edvardsson, Enke, Eriksson, Esquej,
  Eynard~Bontemps, Fabre, Fabrizio, Faigler, Falc{\~a}o, Farr{\`a}s~Casas,
  Federici, Fedorets, Fernique, Figueras, Filippi, Findeisen, Fonti, Fraile,
  Fraser, Fr{\'e}zouls, Gai, Galleti, Garabato, {Garc{\'i}a-Sedano}, Garofalo,
  Garralda, Gavel, Gavras, Gerssen, Geyer, Giacobbe, Gilmore, Girona,
  Giuffrida, Glass, Gomes, Granvik, Gueguen, Guerrier, Guiraud,
  {Guti{\'e}rrez-S{\'a}nchez}, Haigron, Hatzidimitriou, Hauser, Haywood,
  Heiter, Helmi, Heu, Hilger, Hobbs, Hofmann, Holland, Huckle, Hypki, Icardi,
  Jan{\ss}en, {Jevardat de Fombelle}, Jonker, Juh{\'a}sz, Julbe, Karampelas,
  Kewley, Klar, Kochoska, Kohley, Kolenberg, Kontizas, Kontizas, Koposov,
  Kordopatis, {Kostrzewa-Rutkowska}, Koubsky, Lambert, Lanza, Lasne, Lavigne,
  Le~Fustec, {Le Poncin-Lafitte}, Lebreton, Leccia, Leclerc, {Lecoeur-Taibi},
  Lenhardt, Leroux, Liao, Licata, Lindstr{\o}m, Lister, Livanou, Lobel,
  L{\'o}pez, Managau, Mann, Mantelet, Marchal, Marchant, Marconi, Marinoni,
  Marschalk{\'o}, Marshall, Martino, Marton, Mary, Massari, Matijevi{\v c},
  Mazeh, McMillan, Messina, Michalik, Millar, Molina, Molinaro, Moln{\'a}r,
  Montegriffo, Mor, Morbidelli, Morel, Morris, Mulone, Muraveva, Musella,
  Nelemans, Nicastro, Noval, O'Mullane, Ord{\'e}novic,
  {Ord{\'o}{\~n}ez-Blanco}, Osborne, Pagani, Pagano, Pailler, Palacin,
  Palaversa, Panahi, Pawlak, Piersimoni, Pineau, Plachy, Plum, Poggio,
  Poujoulet, Pr{\v s}a, Pulone, Racero, Ragaini, Rambaux, {Ramos-Lerate},
  Regibo, Reyl{\'e}, Riclet, Ripepi, Riva, Rivard, Rixon, Roegiers, Roelens,
  {Romero-G{\'o}mez}, Rowell, Royer, {Ruiz-Dern}, Sadowski,
  Sagrist{\`a}~Sell{\'e}s, Sahlmann, Salgado, Salguero, Sanna, {Santana-Ros},
  Sarasso, Savietto, Schultheis, Sciacca, Segol, Segovia, S{\'e}gransan, Shih,
  Siltala, Silva, Smart, Smith, Solano, Solitro, Sordo, Soria~Nieto, Souchay,
  Spagna, Spoto, Stampa, Steele, Steidelm{\"u}ller, Stephenson, Stoev, Suess,
  Surdej, Szabados, {Szegedi-Elek}, Tapiador, Taris, Tauran, Taylor, Teixeira,
  Terrett, Teyssandier, Thuillot, Titarenko, Torra~Clotet, Turon, Ulla,
  Utrilla, Uzzi, Vaillant, Valentini, Valette, {van Elteren}, Van~Hemelryck,
  {van Leeuwen}, Vaschetto, Vecchiato, Veljanoski, Viala, Vicente, Vogt, {von
  Essen}, Voss, Votruba, Voutsinas, Walmsley, Weiler, Wertz, Wevers,
  Wyrzykowski, Yoldas, {\v Z}erjal, Ziaeepour, Zorec, Zschocke, Zucker,
  Zurbach, \& Zwitter}]{GaiaCollaborationGaiaDataRelease2018}
{Gaia Collaboration}, Brown, A. G.~A., Vallenari, A., {et~al.} 2018, Astronomy
  and Astrophysics, 616, A1

\bibitem[{Gandhi \& Madhusudhan(2019)}]{GandhiNewAvenuesThermal2019}
Gandhi, S., \& Madhusudhan, N. 2019, Monthly Notices of the Royal Astronomical
  Society, 485, 5817

\bibitem[{Gelman \& Rubin(1992)}]{GelmanInferenceIterativeSimulation1992}
Gelman, A., \& Rubin, D.~B. 1992, Statistical Science, 7, 457

\bibitem[{Gibson {et~al.}(2019)Gibson, {de Mooij}, Evans, Merritt, Nikolov,
  Sing, \& Watson}]{GibsonRevisitingPotassiumFeature2019}
Gibson, N.~P., {de Mooij}, E. J.~W., Evans, T.~M., {et~al.} 2019, Monthly
  Notices of the Royal Astronomical Society, 482, 606

\bibitem[{Gibson {et~al.}(2017)Gibson, Nikolov, Sing, Barstow, Evans, Kataria,
  \& Wilson}]{GibsonVLTFORS2Comparative2017}
Gibson, N.~P., Nikolov, N., Sing, D.~K., {et~al.} 2017, Monthly Notices of the
  Royal Astronomical Society, 467, 4591

\bibitem[{Herman {et~al.}(2020)Herman, {de Mooij}, Jayawardhana, \&
  Brogi}]{HermanSearchTiOOptical2020}
Herman, M.~K., {de Mooij}, E. J.~W., Jayawardhana, R., \& Brogi, M. 2020, The
  Astronomical Journal, 160, 93

\bibitem[{Herman {et~al.}(2022)Herman, {de Mooij}, Nugroho, Gibson, \&
  Jayawardhana}]{HermanDaysideFeEmission2022}
Herman, M.~K., {de Mooij}, E. J.~W., Nugroho, S.~K., Gibson, N.~P., \&
  Jayawardhana, R. 2022, The Astronomical Journal, 163, 248

\bibitem[{Hunter(2007)}]{Hunter:2007}
Hunter, J.~D. 2007, Computing in Science \& Engineering, 9, 90

\bibitem[{Kesseli {et~al.}(2020)Kesseli, Snellen, {Alonso-Floriano},
  Molli{\`e}re, \& Serindag}]{KesseliSearchFeHHotJupiter2020}
Kesseli, A.~Y., Snellen, I. A.~G., {Alonso-Floriano}, F.~J., Molli{\`e}re, P.,
  \& Serindag, D.~B. 2020, The Astronomical Journal, 160, 228

\bibitem[{Kirkpatrick {et~al.}(1999)Kirkpatrick, Reid, Liebert, Cutri, Nelson,
  Beichman, Dahn, Monet, Gizis, \&
  Skrutskie}]{KirkpatrickDwarfsCoolerDefinition1999}
Kirkpatrick, J.~D., Reid, I.~N., Liebert, J., {et~al.} 1999, The Astrophysical
  Journal, 519, 802

\bibitem[{Line {et~al.}(2021)Line, Brogi, Bean, Gandhi, Zalesky, Parmentier,
  Smith, Mace, Mansfield, Kempton, Fortney, Shkolnik, Patience, Rauscher,
  D{\'e}sert, \& Wardenier}]{LineSolarSubsolarMetallicity2021}
Line, M.~R., Brogi, M., Bean, J.~L., {et~al.} 2021, Nature, 598, 580

\bibitem[{Lockwood {et~al.}(2014)Lockwood, Johnson, Bender, Carr, Barman,
  Richert, \& Blake}]{LockwoodNearIRDirectDetection2014}
Lockwood, A.~C., Johnson, J.~A., Bender, C.~F., {et~al.} 2014, The
  Astrophysical Journal Letters, 783, L29

\bibitem[{MacDonald \& Lewis(2022)}]{MacDonaldTRIDENTRapid3D2022}
MacDonald, R.~J., \& Lewis, N.~K. 2022, The Astrophysical Journal, 929, 20

\bibitem[{MacDonald \&
  Madhusudhan(2017)}]{MacDonaldSignaturesNitrogenChemistry2017}
MacDonald, R.~J., \& Madhusudhan, N. 2017, The Astrophysical Journal Letters,
  850, L15

\bibitem[{MacDonald \&
  Madhusudhan(2019)}]{MacDonaldMetalrichAtmosphereExoNeptune2019}
---. 2019, Monthly Notices of the Royal Astronomical Society, 486, 1292

\bibitem[{Mandel \& Agol(2002)}]{MandelAnalyticLightCurves2002}
Mandel, K., \& Agol, E. 2002, The Astrophysical Journal, 580, L171

\bibitem[{Mart{\'i}n {et~al.}(1999)Mart{\'i}n, Delfosse, Basri, Goldman,
  Forveille, \& Zapatero~Osorio}]{MartinSpectroscopicClassificationLateM1999}
Mart{\'i}n, E.~L., Delfosse, X., Basri, G., {et~al.} 1999, The Astronomical
  Journal, 118, 2466

\bibitem[{Martioli {et~al.}(2012)Martioli, Teeple, Manset, Devost, Withington,
  Venne, \& Tannock}]{MartioliOpenSourcePipeline2012}
Martioli, E., Teeple, D., Manset, N., {et~al.} 2012, 8451, 84512B

\bibitem[{McGruder {et~al.}(2020)McGruder, {L{\'o}pez-Morales}, Espinoza,
  Rackham, Apai, Jord{\'a}n, Osip, Alam, Bixel, Fortney, Henry, Kirk, Lewis,
  Rodler, \& Weaver}]{McGruderACCESSConfirmationNo2020}
McGruder, C.~D., {L{\'o}pez-Morales}, M., Espinoza, N., {et~al.} 2020, The
  Astronomical Journal, 160, 230

\bibitem[{Merritt {et~al.}(2021)Merritt, Gibson, Nugroho, {de Mooij}, Hooton,
  Lothringer, Matthews, {Mikal-Evans}, Nikolov, Sing, \&
  Watson}]{MerrittInventoryAtomicSpecies2021}
Merritt, S.~R., Gibson, N.~P., Nugroho, S.~K., {et~al.} 2021, Monthly Notices
  of the Royal Astronomical Society, 506, 3853

\bibitem[{{NASA Exoplanet Archive}(2021)}]{NEA13}
{NASA Exoplanet Archive}. 2021, Planetary Systems Composite Parameters, Last Access: 20XX-YY-ZZ,
  NExScI-Caltech/IPAC, \dodoi{10.26133/NEA13}

\bibitem[{Nugroho {et~al.}(2017)Nugroho, Kawahara, Masuda, Hirano, Kotani, \&
  Tajitsu}]{NugrohoHighresolutionSpectroscopicDetection2017}
Nugroho, S.~K., Kawahara, H., Masuda, K., {et~al.} 2017, The Astronomical
  Journal, 154, 221

\bibitem[{Oliphant(2006)}]{oliphant2006guide}
Oliphant, T.~E. 2006 ({Trelgol Publishing USA})

\bibitem[{pandas~development {team}(2020)}]{reback2020pandas}
pandas~development {team}, T. 2020, Zenodo

\bibitem[{Pearson {et~al.}(2014)Pearson, Turner, \&
  Sagan}]{PearsonPhotometricObservationHATP16b2014}
Pearson, K.~A., Turner, J.~D., \& Sagan, T.~G. 2014, New Astronomy, 27, 102

\bibitem[{Rathcke {et~al.}(2021)Rathcke, MacDonald, Barstow, Goyal,
  {Lopez-Morales}, Mendon{\c c}a, {Sanz-Forcada}, Henry, Sing, Alam, Lewis,
  Chubb, Taylor, Nikolov, \& Buchhave}]{RathckeHSTPanCETProgram2021}
Rathcke, A.~D., MacDonald, R.~J., Barstow, J.~K., {et~al.} 2021, The
  Astronomical Journal, 162, 138

\bibitem[{Reid {et~al.}(1995)Reid, Hawley, \&
  Gizis}]{ReidPalomarMSUNearbyStar1995}
Reid, I.~N., Hawley, S.~L., \& Gizis, J.~E. 1995, The Astronomical Journal,
  110, 1838

\bibitem[{Ribas {et~al.}(2023)Ribas, Reiners, Zechmeister, Caballero, Morales,
  Sabotta, Baroch, Amado, Quirrenbach, Abril, Aceituno, {Anglada-Escud{\'e}},
  Azzaro, Barrado, B{\'e}jar, {Ben{\'i}tez de Haro}, Bergond, Bluhm,
  Calvo~Ortega, Cardona~Guill{\'e}n, Chaturvedi, Cifuentes, Colom{\'e}, Cont,
  {Cort{\'e}s-Contreras}, Czesla, {D{\'i}ez-Alonso}, Dreizler, {Duque-Arribas},
  Espinoza, Fern{\'a}ndez, Fuhrmeister, {Galad{\'i}-Enr{\'i}quez},
  {Garc{\'i}a-L{\'o}pez}, {Gonz{\'a}lez-{\'A}lvarez},
  Gonz{\'a}lez~Hern{\'a}ndez, Guenther, {de Guindos}, Hatzes, Henning, Herrero,
  Hintz, Huelmo, Jeffers, Johnson, {de Juan}, Kaminski, Kemmer, Khaimova,
  Khalafinejad, Kossakowski, K{\"u}rster, Labarga, Lafarga, Lalitha,
  Lamp{\'o}n, {Lillo-Box}, Lodieu, L{\'o}pez~Gonz{\'a}lez, {L{\'o}pez-Puertas},
  Luque, Mag{\'a}n, Mancini, Marfil, Mart{\'i}n, {Mart{\'i}n-Ruiz},
  Molaverdikhani, Montes, Nagel, Nortmann, Nowak, Pall{\'e}, Passegger, Pavlov,
  Pedraz, Perdelwitz, Perger, {Ram{\'o}n-Ballesta}, Reffert, Revilla,
  Rodr{\'i}guez, {Rodr{\'i}guez-L{\'o}pez}, Sadegi, S{\'a}nchez~Carrasco,
  {S{\'a}nchez-L{\'o}pez}, {Sanz-Forcada}, Sch{\"a}fer, Schlecker, Schmitt,
  Sch{\"o}fer, Schweitzer, Seifert, Shan, Skrzypinski, Solano, Stahl, Stangret,
  Stock, St{\"u}rmer, Tabernero, {Tal-Or}, Trifonov, Vanaverbeke, Yan, \&
  Zapatero~Osorio}]{RibasCARMENESSearchExoplanets2023}
Ribas, I., Reiners, A., Zechmeister, M., {et~al.} 2023, Astronomy and
  Astrophysics, 670, A139

\bibitem[{Rodler {et~al.}(2013)Rodler, K{\"u}rster, \&
  Barnes}]{RodlerDetectionCOAbsorption2013}
Rodler, F., K{\"u}rster, M., \& Barnes, J.~R. 2013, Monthly Notices of the
  Royal Astronomical Society, 432, 1980

\bibitem[{{S{\'a}nchez-L{\'o}pez} {et~al.}(2022){S{\'a}nchez-L{\'o}pez},
  Landman, Molli{\`e}re, {Casasayas-Barris}, Kesseli, \&
  Snellen}]{Sanchez-LopezSearchingOriginEhrenreich2022}
{S{\'a}nchez-L{\'o}pez}, A., Landman, R., Molli{\`e}re, P., {et~al.} 2022,
  Astronomy and Astrophysics, 661, A78

\bibitem[{Sing {et~al.}(2015)Sing, Wakeford, Showman, Nikolov, Fortney,
  Burrows, Ballester, Deming, Aigrain, D{\'e}sert, Gibson, Henry, Knutson,
  {Lecavelier des Etangs}, Pont, {Vidal-Madjar}, Williamson, \&
  Wilson}]{SingHSTHotJupiterTransmission2015}
Sing, D.~K., Wakeford, H.~R., Showman, A.~P., {et~al.} 2015, Monthly Notices of
  the Royal Astronomical Society, 446, 2428

\bibitem[{Skaf {et~al.}(2020)Skaf, Bieger, Edwards, Changeat, Morvan, Kiefer,
  Blain, Zingales, Poveda, {Al-Refaie}, Baeyens, Gressier, Guilluy, Jaziri,
  {Modirrousta-Galian}, Mugnai, Pluriel, Whiteford, Wright, Yip, Charnay,
  Leconte, Drossart, Tsiaras, Venot, Waldmann, \&
  Beaulieu}]{SkafARESIICharacterizing2020}
Skaf, N., Bieger, M.~F., Edwards, B., {et~al.} 2020, The Astronomical Journal,
  160, 109

\bibitem[{Sotzen {et~al.}(2020)Sotzen, Stevenson, Sing, Kilpatrick, Wakeford,
  Filippazzo, Lewis, H{\"o}rst, {L{\'o}pez-Morales}, Henry, Buchhave,
  Ehrenreich, Fraine, Garc{\'i}a~Mu{\~n}oz, Jayaraman, Lavvas, {Lecavelier des
  Etangs}, Marley, Nikolov, Rathcke, \&
  {Sanz-Forcada}}]{SotzenTransmissionSpectroscopyWASP79b2020}
Sotzen, K.~S., Stevenson, K.~B., Sing, D.~K., {et~al.} 2020, The Astronomical
  Journal, 159, 5

\bibitem[{Spring {et~al.}(2022)Spring, Birkby, Pino, Alonso, Hoyer, Young,
  Coelho, Nespral, \& {L{\'o}pez-Morales}}]{SpringBlackMirrorImpact2022}
Spring, E.~F., Birkby, J.~L., Pino, L., {et~al.} 2022, Astronomy and
  Astrophysics, 659, A121

\bibitem[{Turner {et~al.}(2022)Turner, Flagg, {Ridden-Harper}, \&
  Jayawardhana}]{TurnerCharacterizingWASP4System2022}
Turner, J.~D., Flagg, L., {Ridden-Harper}, A., \& Jayawardhana, R. 2022, The
  Astronomical Journal, 163, 281

\bibitem[{Turner {et~al.}(2021)Turner, {Ridden-Harper}, \&
  Jayawardhana}]{TurnerDecayingOrbitHot2021}
Turner, J.~D., {Ridden-Harper}, A., \& Jayawardhana, R. 2021, The Astronomical
  Journal, 161, 72

\bibitem[{Turner {et~al.}(2017)Turner, Leiter, Biddle, Pearson,
  {Hardegree-Ullman}, Thompson, Teske, Cates, Cook, Berube, Nieberding, Jones,
  Raphael, Wallace, Watson, \&
  Johnson}]{TurnerInvestigatingPhysicalProperties2017}
Turner, J.~D., Leiter, R.~M., Biddle, L.~I., {et~al.} 2017, Monthly Notices of
  the Royal Astronomical Society, 472, 3871

\bibitem[{Van Der~Walt {et~al.}(2011)Van Der~Walt, Colbert, \&
  Varoquaux}]{van2011numpy}
Van Der~Walt, S., Colbert, S.~C., \& Varoquaux, G. 2011, Computing in Science
  \& Engineering, 13, 22

\bibitem[{Virtanen {et~al.}(2020)Virtanen, Gommers, Oliphant, Haberland, Reddy,
  Cournapeau, Burovski, Peterson, Weckesser, Bright, {van der Walt}, Brett,
  Wilson, Millman, Mayorov, Nelson, Jones, Kern, Larson, Carey, Polat, Feng,
  Moore, VanderPlas, Laxalde, Perktold, Cimrman, Henriksen, Quintero, Harris,
  Archibald, Ribeiro, Pedregosa, {van Mulbregt}, \& {SciPy 1.0
  Contributors}}]{2020SciPy-NMeth}
Virtanen, P., Gommers, R., Oliphant, T.~E., {et~al.} 2020, Nature Methods, 17,
  261

\bibitem[{Visscher {et~al.}(2010)Visscher, Lodders, \&
  Fegley}]{VisscherAtmosphericChemistryGiant2010}
Visscher, C., Lodders, K., \& Fegley, Jr., B. 2010, The Astrophysical Journal,
  716, 1060

\bibitem[{Welbanks {et~al.}(2019)Welbanks, Madhusudhan, Allard, Hubeny,
  Spiegelman, \& Leininger}]{WelbanksMassMetallicityTrendsTransiting2019}
Welbanks, L., Madhusudhan, N., Allard, N.~F., {et~al.} 2019, The Astrophysical
  Journal, 887, L20

\bibitem[{Wenger {et~al.}(2000)Wenger, Ochsenbein, Egret, Dubois, Bonnarel,
  Borde, Genova, Jasniewicz, Lalo{\"e}, Lesteven, \&
  Monier}]{WengerSIMBADastronomicaldatabase2000}
Wenger, M., Ochsenbein, F., Egret, D., {et~al.} 2000, Astronomy and
  Astrophysics Supplement Series, 143, 9

\bibitem[{Zacharias {et~al.}(2013)Zacharias, Finch, Girard, Henden, Bartlett,
  Monet, \& Zacharias}]{ZachariasFourthUSNaval2013}
Zacharias, N., Finch, C.~T., Girard, T.~M., {et~al.} 2013, The Astronomical
  Journal, 145, 44

\end{thebibliography}

\appendix 
\restartappendixnumbering

\section{TESS Transit Timings} \label{app:TESS}
The mid-transit time for each TESS transit fit can be found in Table \ref{tb:TESS_mid}. 

\begin{table}[!h]
    \centering
    \begin{tabular}{ccc}
    \hline
Epoch   & T$_{c}$ (BJD$_{TDB}$)  & T$_{c}$ 1$\sigma$ Error (BJD$_{TDB}$) \\
    \hline
785	    &  2458547.4879012	&    0.00066 \\
786	    &  2458550.8946979	&    0.00060 \\
787	    &  2458554.2997179	&    0.00061 \\
788	    &  2458557.7055964	&    0.00068 \\
789	    &  2458561.1115885	&    0.00063 \\
1001    &  2459283.1594180	&    0.00061 \\
1002    &  2459286.5652727	&    0.00070 \\
1003	&  2459289.9714827	&   0.00062 \\
1005	&  2459296.7833610	&   0.00071 \\
1006	&  2459300.1886769	&   0.00065 \\
1007	&  2459303.5956452	&   0.00077 \\
\hline
    \end{tabular}
    \caption{Individual TESS (Sectors 9 and 36) mid-transit times for WASP-31b derived using EXOMOP}
    \label{tb:TESS_mid}
\end{table}

\end{document}